\documentclass[preprint,prc,showpacs,preprintnumbers,amsmath,amssymb,floatfix]{revtex4}
\usepackage{amsmath}
\usepackage{graphicx}% Include figure files
\usepackage{dcolumn}% Align table columns on decimal point
\usepackage{bm}% bold math
\usepackage{ulem} %% for strike-through
\usepackage[usenames]{color}
\usepackage{epstopdf}
\usepackage{epsfig}
\usepackage{float}
\usepackage{subfigure}
\usepackage{multirow}
\usepackage[version=3]{mhchem}

\newcommand{\nc}{\newcommand}       % new command
\nc{\vc}[1] {\mbox{\boldmath $#1$}} % boldmath(vector)
\nc{\del}       {\partial}              % bra state
\nc{\bra}       {\langle}               % bra state
\nc{\ket}       {\rangle}               % ket state
\nc{\bras}[1]   {\langle #1|}           % bra state
\nc{\kets}[1]   {|#1\rangle}            % ket state
\nc{\mapleft}[1]{           % something under arrow
 \smash{\mathop{\,          %
  \hbox to 1.5cm{\rightarrowfill}\, }\limits_{#1}}}
\nc{\beq}     {\begin{eqnarray}} \nc{\eeq}    {\end{eqnarray}}
\nc{\nn}      {\\\nonumber} \nc{\vs}      {\vspace{-0.275cm}}
\nc{\fra}    {\frac{1}{2}}
\nc{\mb}        {\mathbf}

\begin{document}

\preprint{}

\title{Quark mean field model with pion and gluon corrections for $\Lambda,~\Xi^0$ hypernuclei and neutron stars}

\author{Xueyong Xing}
\affiliation{School of Physics, Nankai University, Tianjin 300071,  China}
\author{Jinniu Hu}
\email{hujinniu@nankai.edu.cn}
\affiliation{School of Physics, Nankai University, Tianjin 300071,  China}
\author{Hong Shen}
\affiliation{School of Physics, Nankai University, Tianjin 300071,  China}

\date{\today}% It is always \today, today,
\begin{abstract}
Properties of $\Lambda,~\Xi^0$ hypernuclei and neutron stars are investigated in quark mean field model with pion and gluon corrections.  Firstly, $u, ~d$, and $s$ quarks are confined by relativistic harmonic oscillator potentials to generate the baryons, like nucleon, $\Lambda, ~\Sigma$, and $\Xi$ hyperons. The effects of pion-quark coupling and one-gluon exchange are considered perturbatively.  Then, the baryons interact with each other through exchanging $\sigma, ~\omega,$ and $\rho$ mesons between quarks in hypernuclei and nuclear matter. The strengths of confinement potentials for $u, ~d,$ and $s$ quarks are determined by the masses and radii of free baryons. The coupling constants between the quarks and mesons are fixed by the ground-state properties of several nuclei and single-hyperon potentials at nuclear saturation density, which arises three parameter sets for the coupling constants between mesons and quarks, named QMF-NK1S, QMF-NK2S, and QMF-NK3S.  Compared to the results of previous quark mean field model without pion and gluon corrections, it is found that properties of $\Lambda$ hypernuclei, i.e. the single $\Lambda$ energies, are more consistent with the experimental observables. Properties of $\Xi^0$ hypernuclei are also calculated and compared with the results in previous quark mean field model. With these three parameter sets, the neutron stars containing hyperons are investigated through solving the Tolman-Oppenheimer-Volkoff equation. Maximum masses of neutron stars approach $2.1M_\odot$ with hyperons and corresponding radii are around $13$ km.
\end{abstract}


%Use showkeys class option if keyword

\maketitle

\section{Introduction}

With the energy and density increasing, the strangeness degree of freedom will appear in nuclear physics to generate hypernuclei and strange nuclear matter, even exist in the core region of neutron stars. Many large facilities are attempting to produce a lot of experimental data about the hypernuclei to further investigate the strangeness nuclear physics, such as J-PARC, MAMI, JLab, FAIR, and so on~\cite{feliciello15}. Until now, $\Lambda$ hypernuclei are the most familiar objects in strangeness physics and have rich experimental events, that were largely created in laboratories from $\ce{_\Lambda^{3}H}$ to $\ce{_\Lambda^{208}Pb}$, while it is generally recognized that $\Sigma$ hypernuclei  do not exist except $^4_\Sigma$He.  Furthermore, there are only several observed data on $\Xi$ hypernuclei in $\ce{^{12}C}+\Xi^-$ and $\ce{^{14}N}+\Xi^-$ systems~\cite{aoki93,yamaguchi01,nakazawa15,gogami16}.

On the other hand, the theoretical study of hypernuclei has been in advance on the basis of a number of experiments on hypernuclei~\cite{danysz63,prowse66,aoki91,franklin95,takahashi01,gal10,botta12,tam12}. The light hypernuclei can be described precisely by the \textit{ab initio} calculation with realistic hyperon-nucleon interaction~\cite{hiyama09}.  For the systematic study of both light and heavy hypernuclei, nonrelativistic and covariant density functional theories are generally employed, such as Skyrme Hartree-Fock (SHF) model~\cite{guleria12,li13,schulze13,cui15,zhou16}, relativistic mean field (RMF) model~\cite{sugahara94,shen06,xu12,sun16},  relativistic point-coupling model~\cite{tanimura12}, and so on.

Since the massive neutron stars ($M\sim 2M_\odot$) were observed~\cite{demorest10,antoniadis13,fonseca16}, it was found that a large number of available theoretic models, which can describe the properties of hypernuclei very well, lead to too soft equations of state (EOS) with hyperon to satisfy the astronomic observables. This dilemma is called  hyperon puzzle. To solve such problem, many mechanisms were proposed to introduce an extra repulsion to make the EOS stiffer, such as the inclusion of a more repulsive hyperon-hyperon force~\cite{weissenborn12,oertel15}, inclusion of a three-body hyperon force~\cite{yamamoto13}, and quark star~\cite{demorest10}. 

Most of available nuclear models are based on the understanding of baryons as fundamental particles, however in quantum chromodynamics (QCD) which is regarded as the standard theory to dominate the strong interaction, quarks and gluons are considered as fundamental particles. They  constitute  mesons and baryons such as pion, nucleon, and hyperons. These compound particles combine together to form a nucleus via the baryon-baryon force in terms of the residual part of strong interaction at hadron level. The nuclear many-body system cannot be solved directly from QCD theory until now due to its non-perturbative feature at low energies. Therefore, it is very important to consider nucleon structure in nuclear many-body system from quark degree of freedom.

A few of attempts have been worked out to study the strangeness nuclear physics at quark level. Quark-meson coupling (QMC) model~\cite{panda97,panda99,panda03,panda04,panda12,saito07} and quark mean field (QMF) model~\cite{toki98,shen00,shen02,hu14a,hu14b,wang01,wang02,wang03,wang04a,wang04b,wang05} are two most successful schemes, where quarks are confined to form a baryon by MIT bag model and confinement potential model, respectively. The baryons in hypernuclei and strange nuclear matter interact with each other via exchanging $\sigma, ~\omega,$ and $\rho$ mesons between the quarks in different baryons.  The properties of baryons will be changed due to the influence of surrounding baryons, which can explain the medium modification of the nucleon structure function (EMC effect)~\cite{aubert83}. However, in these two models, two essentials of QCD theory were not included, chiral symmetry and gluon. At mean field level, the pion contribution is zero, therefore recently many works started to consider the pion effect in QMC model through the exchange term~\cite{krein99,stone07,whittenbury14}. Furthermore, Nagai \textit{et al.} developed the QMC model to include the gluon and pion effects by using the cloudy bag model (CBM)~\cite{nagai08}.  

Several years ago, Mishra \textit{et al.} attempted to overcome such difficulties via calculating the contributions of quark-pion and quark-gluon interactions by one-pion and one-gluon exchanging terms, within lowest-order perturbation theory in a modified quark meson coupling (MQMC) model~\cite{barik13,mishra15} and applied such model to the investigation of strange nuclear matter and neutron stars~\cite{mishra16}. They obtained the massive neutron stars, whose masses are around $2M_\odot$ after fitting the hyperon coupling constants by the empirical $\Lambda-,~\Sigma-$ and $\Xi-$nucleon single-particle potentials at nuclear saturation density. However, in their works, both the coupling constants between the mesons and nucleons and those between mesons and hyperons were determined by the empirical properties of strange nuclear matter at saturation density. Actually, these coupling constants should be also constrained by the properties of normal nuclei and hypernuclei. With coupling constants given by Mishra \textit{et al.}, the properties of nuclei and hypernuclei were not consistent with the experimental data. 

In the past few years, we studied the structures of nuclei and neutron stars in QMF model with pion and gluon effects~\cite{xing16} following the scheme of Mishra \textit{et al.} It was found that the effects of pion and gluon could improve the description of QMF model on nuclear many-body system containing $u$ and $d$ quarks. In present work, this framework will be extended to include the strangeness degree of freedom to study properties of $\Lambda,~\Xi^0$ hypernuclei and neutron stars with hyperons and compare with our previous studies on strangeness system in the QMF model without pion and gluon effects~\cite{hu14a,hu14b}, where the hyperon binding energies of $\Lambda$ and $\Xi^0$ hypernuclei and properties of neutron stars were calculated.    

The paper is written as follows. In Sec. II, we briefly derive the formulas of QMF model with pion and gluon corrections including strangeness degree of freedom. In Sec. III, the new parameter sets of QMF model with pion and gluon corrections for hypernuclei will be determined. The properties of $\Lambda$ and $\Xi^0$ hypernuclei with the new parameter sets will be shown. The neutron stars with hyperons will be also investigated. A summary is given in Sec. IV.

%---------------------------------
\section{Quark mean field model with pion and gluon corrections for hypernuclei and neutron stars}
The analytical confinement potential for quarks cannot be obtained from QCD theory directly. Many phenomenological confinement potentials have been proposed, where the polynomial forms are widely used. A harmonic oscillator potential mixing scalar and vector Lorentz structures, $U(r)$, is adopted in this work, where the Dirac equation can be solved analytically~\cite{xing16}:
\beq\label{1}
U(r)=\frac{1}{2}(1+\gamma^0)(a_qr^2+V_q),
\eeq
where  $q$ denotes $u$, $d$, or $s$. When the effect of nuclear medium is considered, the quark field
$\psi_q(\vec{r})$ satisfies the following Dirac equation:
\beq\label{3}
&&[\gamma^{0}(\epsilon_{q}-g^q_\omega\omega-\tau_{3q}g^q_\rho\rho)
-\vec{\gamma}\cdot\vec{p}-(m_{q}
-g^q_\sigma\sigma)-U(r)]\psi_{q}(\vec{r})=0,
\eeq
where $\sigma$, $\omega$, and $\rho$ are the classical meson fields, which describe the exchanging interaction between quarks. $~g^q_\sigma, ~g^q_\omega$, and $g^q_\rho$ are the coupling constants of $\sigma, ~\omega$, and $\rho$ mesons with quarks, respectively. $\tau_{3q}$ is the third component of isospin matrix and $m_q$ is the bare quark mass. Now we can define the following quantities for convenience:
\beq\label{4}
\epsilon^{\prime}_q&=&\epsilon_q^*-V_q/2,\nn
m^{\prime}_q&=&m_q^*+V_q/2,
\eeq
where the effective quark energy is given by $\epsilon_q^*=\epsilon_{q}-g^q_\omega\omega-\tau_{3q}g^q_\rho\rho$ and the effective quark mass by $m_q^*=m_{q}-g^q_\sigma\sigma$~\cite{xing16}. We also introduce $\lambda_q$ and $r_{0q}$ as
\beq\label{5}
\lambda_q&=&\epsilon^{\prime}_q+m^{\prime}_q,\nn
r_{0q}&=&(a_q\lambda_q)^{-\frac{1}{4}}.
\eeq
The baryon mass in nuclear medium can be expressed as the binding energy of three quarks named zeroth-order term, after solving the Dirac equation (\ref{3}), formally
\beq\label{6}
E_B^{*0}=\sum_q\epsilon^*_q.
\eeq
The quarks are simply confined in a central confinement potential. Three corrections will be taken into account in the zeroth-order baryon mass in nuclear medium, including the center-of-mass correction $\epsilon_{\rm c.m.}$, the pion correction $\delta M_{B}^\pi$, and the gluon correction $(\Delta E_B)_g$. The pion correction is caused by the chiral symmetry of QCD theory and the gluon correction by the short-range  exchange interaction of quarks. The center-of-mass correction can be expressed by~\cite{xing16} 
\beq\label{7}
\epsilon_{\rm c.m.}=\bras{B}\mathcal{H}_{\rm c.m.}\kets{B},
\eeq
where $\mathcal{H}_{\rm c.m.}$ is the center-of-mass Hamiltonian density and $\kets{B}$ is the baryon state. When the baryon wave function is constructed by the quark wave functions, the center-of-mass correction comes out as~\cite{mishra16} 
\beq
\epsilon_{\rm c.m.}=e_{\rm c.m.}^{(1)}+e_{\rm c.m.}^{(2)},
\eeq
where
\beq
e_{\rm c.m.}^{(1)}&=&\sum_{i=1}^3{\left[\frac{m_{q_i}}{\sum_{k=1}^3 m_{q_k}}\frac{6}
{r_{0q_i}^2(3\epsilon'_{q_i}+m'_{q_i})}\right]},\nn
e_{\rm c.m.}^{(2)}
&=&\frac{1}{2}\bigg[\frac{2}{\sum_k m_{q_k}}
\sum_ia_im_i\langle r_i^2\rangle
+\frac{2}{\sum_k m_{q_k}}\sum_ia_im_i\langle \gamma^0(i)r_i^2\rangle
-\frac{3}{(\sum_k m_{q_k})^2}\sum_ia_im_i^2\langle r_i^2\rangle\nonumber\nn
&&-\frac{1}{(\sum_k m_{q_k})^2}\sum_i\langle \gamma^0(1)a_im_i^2r_i^2\rangle-
\frac{1}{(\sum_k m_{q_k})^2}\sum_i\langle \gamma^0(2)a_im_i^2r_i^2\rangle\notag\nn
&&-\frac{1}{(\sum_k m_{q_k})^2}\sum_i\langle \gamma^0(3)a_im_i^2r_i^2\rangle\bigg].
\eeq
Here, the different expectation values related to radii are listed as follows:
\beq
\langle r_i^2\rangle&=&\frac{(11\epsilon_{qi}'+ m_{qi}')r^2_{0qi}}
{2(3\epsilon_{qi}'+ m_{qi}')},\nn
\langle\gamma^0(i)r_i^2\rangle&=&\frac{(\epsilon_{qi}'+ 11 m_{qi}')r^2_{0qi}}
{2(3\epsilon_{qi}'+ m_{qi}')},\nn
\langle\gamma^0(i)r_j^2\rangle_{i\neq j}&=&\frac{(\epsilon_{qi}'+ 3 m_{qi}')
\langle r^2_j\rangle}{3\epsilon_{qi}'+ m_{qi}'}.
\eeq

In QMF model, the constituent quark masses are obtained due to spontaneous chiral symmetry breaking. It is then natural to generate nearly zero mass pions as the Nambu-Goldstone bosons. Their coupling to the constituent quarks is provided by the chiral symmetry. In order to treat the chiral symmetry properly in the baryon, an elementary pion field is introduced in present model.
The pionic self-energy correction to the nucleon mass becomes
\beq\label{}
\delta M^{\pi}_{N}=-\frac{171}{25}I_{\pi}f_{NN\pi}^{2},
\eeq
where
\beq\label{11}
I_{\pi}=\frac{1}{\pi m_{\pi}^{2}}\int_0^\infty dk\frac{k^{4}u^{2}(k)}{w_{k}^{2}},
\eeq
with the axial-vector nucleon form factor
\beq\label{12}
u(k)
=\left[
1-\frac{3}{2}\frac{k^{2}}{\lambda_{u}(5\epsilon^{\prime}_{u}+7m^{\prime}_{u})}
\right]
e^{-\frac{1}{4}r^{2}_{0u}k^2}
\eeq
and $f_{NN\pi}$ can be obtained from the Goldberg-Triemann relation by using the axial-vector coupling-constant value $g_A$ in this model. The pionic corrections for $\Lambda,~\Sigma,$ and $\Xi$ hyperons become
\beq
\delta M_{\Lambda}^{\pi}&=&-{\frac{108}{25}}f_{NN\pi}^2I_{\pi},\nn
\delta M_{\Sigma}^{\pi}&=&-{\frac{12}{5}}f_{NN\pi}^2I_{\pi},\nn
\delta M_{\Xi}^{\pi}&=&-{\frac{27}{25}}f_{NN\pi}^2I_{\pi}.
\eeq

The one-gluon exchange contribution to the baryon mass is separated into two parts as
\beq\label{}
(\Delta E_{B})_{g}=(\Delta E_{B})_{g}^{E}+(\Delta E_{B})_{g}^{M},
\eeq
where $(\Delta E_{B})_{g}^{E}$ is the color-electric contribution
\beq\label{19}
(\Delta E_{B})_{g}^{E}=\frac{1}{8\pi}\sum\limits_{i,j}\sum\limits_{a=1}^{8}\int\frac{d^{3}r_{i}d^{3}r_{j}}
{|\vec{r}_{i}-\vec{r}_{j}|}\bras{B}J^{0a}_{i}(\vec{r}_{i})J^{0a}_{j}(\vec{r}_{j})\kets{B},
\eeq
and $(\Delta E_{B})_{g}^{M}$ the color-magnetic contribution
\beq\label{20}
(\Delta E_{B})_{g}^{M}=-\frac{1}{8\pi}\sum\limits_{i,j}\sum\limits_{a=1}^{8}\int\frac{d^{3}r_{i}d^{3}r_{j}}
{|\vec{r}_{i}-\vec{r}_{j}|}\bras{B}\vec{J}^{a}_{i}(\vec{r}_{i})\cdot\vec{J}^{a}_{j}(\vec{r}_{j})\kets{B}.
\eeq
Here
\beq\label{}
J^{\mu a}_{i}(x)=g_{c}\bar{\psi}_{q}(x)\gamma^{\mu}\lambda^{a}_{i}\psi_{q}(x)
\eeq
is the $i$th quark color current density, where $\lambda_i^a$ are the usual Gell-Mann SU(3) matrices and
$\alpha_{c}=g_{c}^{2}/4\pi$. Then Eqs. (\ref{19}) and (\ref{20}) can be written as
\beq
(\Delta E_B)_g^E={\alpha_c}(b_{uu}I_{uu}^E+b_{us}I_{us}^E+b_{ss}I_{ss}^E),
\eeq
and
\beq
(\Delta E_B)_g^M={\alpha_c}(a_{uu}I_{uu}^M+a_{us}I_{us}^M+a_{ss}I_{ss}^M),
\eeq
where $a_{ij}$ and $b_{ij}$ are the numerical coefficients depending on each baryon and are given in Table \ref{tab1}. The quantities $I_{ij}^{E}$ and $I_{ij}^{M}$ are given in the following equations:
\beq
I_{ij}^{E}&=&\frac{16}{3{\sqrt \pi}}\frac{1}{R_{ij}}\left[1-
\frac{\alpha_i+\alpha_j}{R_{ij}^2}+\frac{3\alpha_i\alpha_j}{R_{ij}^4}
\right],\nn
I_{ij}^{M}&=&\frac{256}{9{\sqrt \pi}}\frac{1}{R_{ij}^3}\frac{1}{(3\epsilon_i^{'}
+m_{i}^{'})}\frac{1}{(3\epsilon_j^{'}+m_{j}^{'})},
\eeq
with
\beq
R_{ij}^{2}&=&3\left[\frac{1}{({\epsilon_i^{'}}^2-{m_i^{'}}^2)}+
\frac{1}{({\epsilon_j^{'}}^2-{m_j^{'}}^2)}\right],
\nn
\alpha_i&=&\frac{1}{ (\epsilon_i^{'}+m_i^{'})(3\epsilon_i^{'}+m_{i}^{'})}.
\eeq
\begin{table}[H]%\footnotesize%\scriptsize%
\centering
\caption{The coefficients $a_{ij}$ and $b_{ij}$ used in the calculation of the color-electric and and color-magnetic energy contributions due to one-gluon exchange for different baryons.}
\label{tab1}
\begin{tabular}{l c c c c c c}
\hline
\hline
Baryon     & $a_{uu}$ & $a_{us}$ & $a_{ss}$ & $b_{uu}$ & $b_{us}$ & $b_{ss}$\\
\hline
$N$        & $-3$&  0  & 0 & 0 &  0  & 0\\
$\Lambda$  & $-3$&  0  & 0 & 1 & $-2$& 1\\
$\Sigma$   &  1  & $-4$& 0 & 1 & $-2$& 1\\
$\Xi$      &  0  & $-4$& 1 & 1 & $-2$& 1\\
		
\hline
\hline
\end{tabular}
\end{table}

The detailed forms of color-electric and color-magnetic contributions can be found in Ref. \cite{mishra16}.  Finally, taking into account all above energy corrections, the mass of baryon in nuclear medium becomes
\beq\label{mass}
M_B^*=E_B^{*0}-\epsilon_{\rm c.m.}+\delta M_B^\pi+(\Delta E_B)^E_g+
(\Delta E_B)^M_g.
\eeq

Next we would like to connect the baryon in the medium with nuclear objects, such as $\Lambda$ and $\Xi^0$ hypernuclei. A single hypernucleus is treated as a system of many nucleons and one hyperon which interact through exchanging $\sigma$, $\omega$, and $\rho$ mesons. The QMF Lagrangian in the mean-field approximation can be written as \cite{toki98,shen00,shen02,hu14a,hu14b},
\beq
{\cal L}_{\rm QMF}
&=&
\bar\psi\left[ i\gamma_\mu\partial^\mu-M_N^*
-g_\omega \omega \gamma^0
-g_\rho \rho \tau_3\gamma^0
-e\frac{(1-\tau_3)}{2} A \gamma^0
\right]\psi\nn
&&
+\bar\psi_H
\left[i\gamma_\mu\partial^\mu-M_H^*
-g_\omega^H\omega\gamma^0
+\frac{f^H_\omega}{2M_H}\sigma^{0i}\partial_i\omega
\right]\psi_H\nn
&&
-\frac{1}{2} (\bigtriangledown\sigma)^2
-\frac{1}{2} m_\sigma^2\sigma^2
-\frac{1}{3} g_2\sigma^3
-\frac{1}{4} g_3\sigma^4\nn
&&
+\frac{1}{2} (\bigtriangledown\omega)^2
+\frac{1}{2} m_\omega^2\omega^2
+\frac{1}{4} c_3\omega^4\nn
&&
+\frac{1}{2} (\bigtriangledown\rho)^2
+\frac{1}{2} m_\rho^2\rho^2
+\frac{1}{2}(\bigtriangledown A)^2,
\eeq
where $H$ denotes $\Lambda$ or $\Xi^0$ hyperon and the effective masses of baryons, $M^*_N$ and $M^*_H$, are generated from quark model, Eq.~(\ref{mass}).  These effective baryon masses are actually related to scalar mesons in RMF model.  Furthermore, we should emphasize that the nonlinear terms of $\sigma$ and $\omega$ are included additionally in present work comparing with the Lagrangian of MQMC model~\cite{mishra15,mishra16}, since these terms can largely improve the descriptions on properties of finite nuclei as shown in our previous work~\cite{xing16}. The tensor coupling between $\omega$ meson and baryons, $\frac{f^H_\omega}{2M_H}\sigma^{0i}\partial_i\omega$, can improve the description of small spin-orbit splittings of hypernuclei~\cite{shen06,sugahara94}.

The equations of motion of baryons and mesons are obtained by using the Euler-Lagrange equation. Dirac equations for nucleons and hyperons have the following form:
\beq
&&\left[i\gamma_{\mu}\partial^{\mu}-M_N^*
-g_\omega\omega\gamma^0
-g_\rho\rho\tau_3\gamma^0\
-e\frac{(1-\tau_3)}{2}A\gamma^0
\right]\psi
=0,\nn
&&\left[i\gamma_{\mu}\partial^{\mu}-M_H^*
-g_\omega^H\omega\gamma^0
+\frac{f^H_\omega}{2M_H}\sigma^{0i}\partial_i\omega
\right]\psi_H
=0.
\eeq
The equations of motion for mesons are given by
\beq
&&\Delta\sigma-m_\sigma^2\sigma-g_2\sigma^2-g_3\sigma^3
=\frac{\partial M_N^*}{\partial\sigma}
\langle\bar\psi\psi\rangle
+\frac{\partial M_H^*}{\partial\sigma}
\langle\bar\psi_H\psi_H\rangle,\nn
&&\Delta\omega-m_\omega^2\omega-c_3 \omega^3=
-g_\omega\langle\bar\psi\gamma^0\psi\rangle
-g_\omega^H
\langle\bar\psi_H\gamma^0\psi_H\rangle
+\frac{f_\omega^H}{2M_H}
\partial_i\langle\bar\psi_H\sigma^{0i}
\psi_H\rangle,\nn
&&\Delta\rho-m_\rho^2\rho=
-g_\rho\langle\bar\psi\tau_3\gamma^0\psi\rangle,\nn
&&\Delta A=
-e\langle\bar\psi\frac{(1-\tau_3)}{2}\gamma^0\psi\rangle.
\eeq
Here, the coupling constants between $\omega, ~\rho$ mesons and nucleons, $g_\omega$ and $g_\rho$, are generated from quark counting rules, $g_\omega=3g^q_\omega$ and $g_\rho=g^q_\rho$, while those between $\omega$ mesons and hyperons, $g^H_\omega$ and  $f^H_\omega$ will be determined by the properties of hypernuclei and strange nuclear matter at nuclear saturation density. Above equations of motion of baryons and mesons can be solved self-consistently with numerical method. From the single-particle energies of nucleons and hyperon, the total energy of whole hypernucleus can be obtained with mean field approximation.

In strange nuclear matter including $\Lambda,~\Sigma,$ and $\Xi$  hyperons, the gradient terms in the equations of motion of mesons would disappear. The energy density and pressure are generated from the energy-momentum tensor related to QMF Lagrangian. In neutron stars, there are not only baryons but also leptons, such as, electrons and muons. The neutron star matter satisfies electric neutrality and $\beta$ equilibrium. In such case, the EOS can be solved and taken into Tolman-Oppenheimer-Volkoff (TOV) equation~\cite{oppenheimer39,tolman39} to get properties of neutron stars. The detailed formulas can be found in our previous work about QMF model on neutron star~\cite{hu14b}.   

\section{Results and discussion}
\subsection{Properties of baryons}
Firstly, the strengths of quark confinement potentials for $u,~d,$ and $s$ quarks should be determined. In present work, there are two free parameters, $a_q$ and $V_q$, in the confinement potential for each flavor quark.  The differences of properties  between $u$ and $d$ quarks are very small, therefore, they are treat equally in this work. For $s$ quark, SU(3) symmetry is broken, where $a_s$ and $V_s$ are distinguished from $a_u$ and $V_u$. In QMF model, the quarks are regarded as the constituent ones, whose masses are around $300$ MeV for $u$ and $d$ quarks. In recent lattice QCD calculations \cite{parappilly06,burgio12}, the value of constituent quark mass was suggested to be around $250-350$ MeV. To discuss the influences of quark mass, we take the masses of $u$ and $d$ quarks to be $250, ~300$, and $350$ MeV in three parameter sets. The corresponding $a_u$ and $V_u$ are fixed by the mass and radius of free nucleon, which were already given in our previous work~\cite{xing16}. The $s$ quark mass and two coefficients, $a_s$ and $V_s$, in $s$ quark confinement potential are obtained by fitting the free masses of $\Lambda, ~\Sigma^0,$ and $\Xi^0$ hyperons~\cite{patrignani16}  through least-squares method. 

These parameters are listed in Table \ref{tab2}. For convenience, the first parameter set ($m_u=250$ MeV) in Table \ref{tab2} is named as set A, the second ($m_u=300$ MeV) as set B and the third ($m_u=350$ MeV)  as set C. Here, we should emphasize that in the work of Mishra \textit{et al.}~\cite{mishra16}, each baryon corresponds to one $V_q$ value, while in present work, the confinement potentials of $s$ quark are adopted as uniform strength  in $\Lambda, ~\Sigma,$ and $\Xi$ hyperons.  The differences of their masses are generated by the  pion and gluon corrections.

\begin{table}[htb]%\footnotesize%\scriptsize%
\centering
\caption{The potential parameters $(a_q,V_q)$ obtained for the quark mass $m_u=250$ MeV, $m_s=330$ MeV, the quark mass $m_u=300$ MeV, $m_s=380$ MeV, and the quark mass $m_u=350$ MeV, $m_s=430$ MeV. }
\label{tab2}
\begin{tabular}{c c c c c c c}
\hline
\hline
    &$m_u$ (MeV)&$V_u$  (MeV)    &$a_u$ (fm$^{-3}$)       &$m_s$ (MeV) &$V_s$ (MeV)    &$a_s$ (fm$^{-3}$)     \\
\hline

set A&250  &$-24.286601$ &0.579450       &330   &101.78180&0.097317  \\

set B&300  &$-62.257187$ &0.534296       &380   &54.548210&0.087243  \\

set C&350  &$-102.041575$&0.495596       &430   &6.802695 &0.079534  \\

\hline
\hline
\end{tabular}
\end{table}

In Table \ref{tab3},  the masses of three charge neutrality hyperons ($\Lambda,~\Sigma^0$, and $\Xi^0$) in free space with set A, set B, and set C are compared with the latest experimental data~\cite{patrignani16}, respectively. The contributions from center-of-mass, pionic, and gluonic corrections to the hyperon masses in free space are also shown. Under the constraint of $s$ quark mass, whose value should be larger than those of $u,~d$ quarks, the hyperon masses in theoretical calculation do not reproduce the experimental observables~\cite{patrignani16} from the particle data group completely. There are several MeV differences between the theoretical prediction and experimental data. The $1\%$ errors are not able to influence the further calculations and discussions.
\begin{table}[htb]%\footnotesize%\scriptsize%
\centering
\caption{The masses of three hyperons ($\Lambda$, $\Sigma^0$, and $\Xi^0$) in free space from set A, set B, and set C, compared with the experimental data and the contributions of center-of-mass, pionic, and gluonic corrections to the hyperon masses respectively (The units of all quantities are MeV). }
\label{tab3}
\begin{tabular}{c c c c c c c c c}
\hline
\hline
    &Baryon&$E_B^0$&$\epsilon_{\rm c.m.}$&$\delta M_B^\pi$&$(\Delta E_B)_g$      &$M^\text{Theo.}_B$&$M^\text{Exp.}_B$\\
\hline

         &$\Lambda$&$1446.340$&$231.975$&$-65.172$&$-24.390$&$1124.803$&$1115.683\pm0.006$\\

set A&${\Sigma^0}$ &$1446.340$&$231.975$&$-36.207$&$10.515$ &$1188.673$&$1192.642\pm0.024$\\

         &$\Xi^0$           &$1504.254$ &$175.047$&$-16.293$&$-1.289$ &$1311.625$&$1314.86\pm0.20$\\

\hline

         &$\Lambda$&$1433.489$&$220.692$&$-69.277$&$-18.313$&$1125.207$&$1115.683\pm0.006$\\

set B&${\Sigma^0}$ &$1433.489$&$220.692$&$-38.487$&$13.753$&$1188.063$&$1192.642\pm0.024$\\

         &$\Xi^0$            &$1491.611$  &$165.564$&$-17.319$  &$2.979$  &$1311.707$&$1314.86\pm0.20$\\

\hline

         &$\Lambda$&$1421.908$&$210.233$&$-72.829$&$-13.170$&$1125.676$&$1115.683\pm0.006$\\

set C&${\Sigma^0}$ &$1421.908$&$210.233$&$-40.461$&$16.203$&$1187.417$&$1192.642\pm0.024$\\

         &$\Xi^0$            &$1480.703$&$157.102$&$-18.207$ &$6.377$   &$1311.771$&$1314.86\pm0.20$\\

\hline
\hline
\end{tabular}
\end{table}

\begin{figure}[htb]
\centering
\includegraphics[width=12cm]{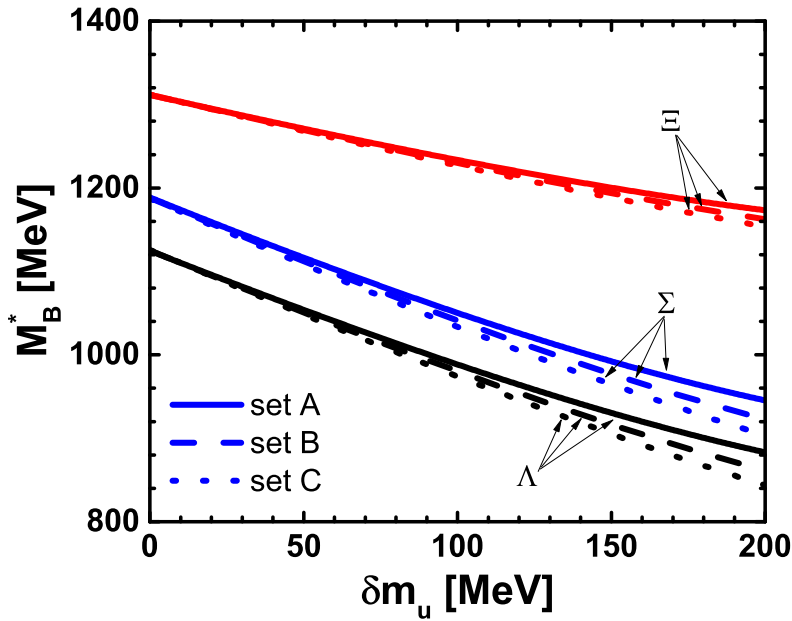}
\caption{The effective masses of three hyperons ($\Lambda,~\Sigma$, and $\Xi$) with different parameter sets [set A (solid curve), set B (dashed curve), and set C (dotted curve)] as functions of $u$ quark mass correction.}
\label{fig1}
\end{figure}

The behaviors of one baryon in nuclear medium will be influenced by the surrounding particles, therefore, its effective mass  $M_B^*$ is able to change with density increasing. In QMF model, medium effect is included via effective quark mass generated by $\sigma$ meson. Finally, the effective baryon masses are the functions of  quark mass corrections  $\delta m_q=m_q-m_q^*=g^q_\sigma\sigma$. The $\sigma$ meson did  not contain the strangeness flavor, so that the coupling constant between $s$ quark and $\sigma$ meson is zero, i.e., $g_\sigma^s=g_\omega^s=0$. All effective masses of baryons are only affected by the $u,~d$ quarks. In Fig.~\ref{fig1}, the effective masses of three hyperons ($\Lambda$, $\Sigma$, and $\Xi$) for different parameter sets (set A, set B, and set C) are given as functions of $u$ quark mass correction.

In free space ($\delta m_u=0$), their effective masses actually correspond to the masses of free hyperons. With $\delta m_u$ increasing, the effective hyperon masses will be reduced in terms of the effect of surrounding baryons. At small quark mass correction, the effective masses are almost the same for different parameter sets. With the quark mass correction $\delta m_u$ increasing, the difference among set A, B, and C becomes obvious for $\Lambda$ and $\Sigma$ hyperons. Since there is only one $u$ quark component in $\Xi$ hyperon, the influences from different parameter sets are very small.

\subsection{Properties of hypernuclei}

Once the relation between the effective baryon masses and quark mass corrections is obtained, the next step is to determine the coupling constants between quarks and mesons: $g_\sigma^u, ~g_\omega, ~g_\omega^\Lambda, ~g_\omega^\Xi, ~f_\omega^H, ~g_\rho$, and the parameters in nonlinear terms of $\sigma$ and $\omega$ mesons, $g_2,~g_3$, and $c_3$. In this work, the meson masses are taken as $m_\sigma=550$ MeV, $m_\omega=783$ MeV, and $m_\rho=763$ MeV. $g_\sigma^u$, $g_\omega$, $g_\rho$, $g_2$, $g_3$, and $c_3$  about the normal nuclei have been obtained by fitting the properties of finite nuclei~\cite{xing16}. 

In the present calculation, we adopt the quark model value of the tensor coupling between $\omega$ and hyperons~\cite{shen06,sugahara94}, $f_\omega^H=-g_\omega^H$, which is important to produce small spin-orbit spitting of hypernuclei. Only the coupling constants between $\omega$ and hyperons need to be conformed. They will be decided via the magnitudes of single hyperon potentials at nuclear saturation density in nuclear matter. The single $\Lambda$ and $\Xi$ potentials in nuclear matter are fixed as $U_\Lambda=-30$ MeV and $U_\Xi=-12$ MeV at nuclear saturation density respectively, following the existing experimental data about $\Lambda$ and $\Xi$ hypernuclei. Based on these choices, the single $\Lambda$ and $\Xi$ potentials in Pb hypernuclei are also calculated to check the validity of $g_\omega^H$. Finally, we obtain three parameter sets about vector coupling constants corresponding to $u$ quark masses $m_u=250$ MeV, $m_u=300$ MeV, and $m_u=350$ MeV and these parameter sets are listed Table \ref{tab4}. For convenience, we name the first parameter set ($m_u=250$ MeV) in Table \ref{tab4} as QMF-NK1S, the second one ($m_u=300$ MeV) as QMF-NK2S, and the third one ($m_u=350$ MeV) as QMF-NK3S. 
\begin{table}[htb]%\footnotesize%\scriptsize%
\centering
\caption{The parameters for quarks and hadrons are listed. The first parameter set corresponding to $m_u=250$ MeV is named QMF-NK1S, the second for $m_u=300$ MeV named QMF-NK2S, and the third for $m_u=350$ MeV named QMF-NK3S.}
\label{tab4}
%\scalebox{1}{
\begin{tabular}{l c c c c c c c c c c}
\hline
\hline
Model&$m_u$&$g_{\sigma}^{u}$&$g_{\omega}$&$g_\omega^\Lambda$&$g_\omega^\Xi$&$g_{\rho}$&$g_2$&$g_3$&$c_3$\\
     &(MeV)&                &            &                  &              &          &$(\rm fm^{-1})$&      &     \\
\hline

QMF-NK1S&250&5.15871      &11.54726    &$0.8258g_\omega$&$0.4965g_\omega$&3.79601   &$-3.52737$   &$-78.52006$  &305.00240\\

QMF-NK2S&300&5.09346    &12.30084    &$0.8134g_\omega$&$0.4800g_\omega$&4.04190   &$-3.42813$   &$-57.68387$  &249.05654\\

QMF-NK3S&350&5.01631      &12.83898    &$0.8040g_\omega$&$0.4681g_\omega$&4.10772   &$-3.29969$   &$-39.87981$  &221.68240\\

\hline
\hline
\end{tabular}%}
\end{table}

The ratios of $g^H_\omega/g_\omega$ for $\Lambda$ and $\Xi $ hyperons in QMF-NK1S, QMF-NK2S, and QMF-NK3S do not satisfy the suggestions from simple  quark counting rules as $2/3$ and $1/3$ used in our previous work~\cite{hu14a,hu14b}. It is because that the cubic term of $\sigma$ meson and biquadratic term of $\omega$ term are included in present work, which generate larger values of $g^u_\sigma$ and $g_\omega$. The corresponding $g^H_\omega$ becomes larger to provide more repulsive vector potential.

\begin{figure}[htb]
	\centering
	\includegraphics[width=15cm]{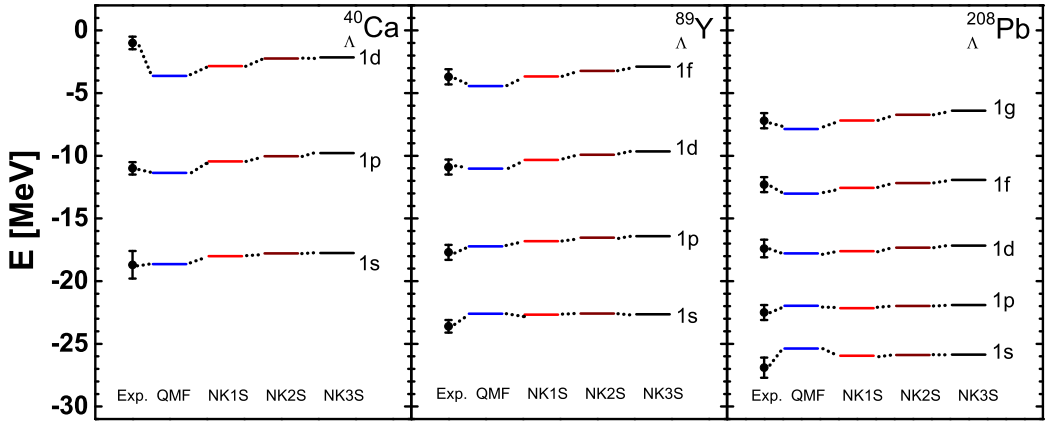}
	\caption{The results of theoretical calculation for energy levels of $\Lambda$ hyperon for $\ce{_\Lambda^{40}Ca}$, $\ce{_\Lambda^{89}Y}$, and $\ce{_\Lambda^{208}Pb}$ by QMF-NK1S, QMF-NK2S, and QMF-NK3S, compared with the experimental data and the results in previous QMF model without pion and gluon effects.}
	\label{lame}
\end{figure}

In Fig. \ref{lame}, the energy levels given by theoretical calculation for $\Lambda$ hyperon in three single $\Lambda$ hypernuclei, $\ce{_\Lambda^{40}Ca}$, $\ce{_\Lambda^{89}Y}$, and $\ce{_\Lambda^{208}Pb}$, within QMF-NK1S, QMF-NK2S, and QMF-NK3S parameter sets are compared with the experimental data~\cite{gal16}. Here, we should make a statement that the single $\Lambda$ binding energies listed in recent reviews article~\cite{gal16} are not same as the well-known data summarized by Hashimoto and Tamura~\cite{hashimoto06}, since they revised these data with latest experimental information of light hypernuclei in the past few years. The results from our previous QMF calculation without pion and gluon corrections~\cite{hu14a} are also given for comparison. We can find that the energy level of $1d$ state in $\ce{_\Lambda^{40}Ca}$ are largely improved in present model compared with that from the QMF model without pion and gluon corrections. In $\ce{_\Lambda^{89}Y}$ and $\ce{_\Lambda^{208}Pb}$, all energy levels are polished up in present calculations to accord with experiment data better.  Generally speaking, the $\Lambda$ energy levels as a whole in QMF-NK3S set are larger than those in QMF-NK1S. It is related to the coupling constants without strangeness degree of freedom. For example, the binding energy of $^{208}$Pb from QMF-NK3 is smaller than that from QMF-NK1 as shown in Ref.~\cite{xing16}.

\begin{figure}[htb]
	\centering
	\includegraphics[width=15cm]{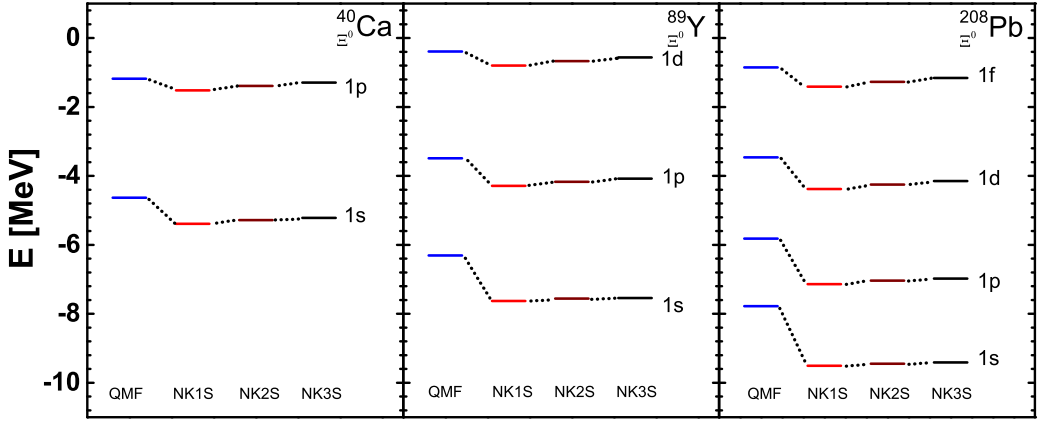}
	\caption{The results of theoretical calculation for energy levels of $\Xi^0$ hyperon for $\ce{_{\Xi^0}^{40}Ca}$, $\ce{_{\Xi^0}^{89}Y}$, and $\ce{_{\Xi^0}^{208}Pb}$ by QMF-NK1S, QMF-NK2S, and QMF-NK3S, compared with the results in our previous QMF model without pion and gluon corrections.}
	\label{xie}
\end{figure}

 Encouraged by the good agreements of $\Lambda$ hypernuclei data in our present model, we start to calculate the energy levels of  $\Xi^0$ hypernuclei in the same framework, to serve as a reference for the future experiments. The single-particle energy levels of $\Xi^0$ for $\ce{^{40}_{\Xi^0}Ca}$, $\ce{^{89}_{\Xi^0}Y}$,and $\ce{^{208}_{\Xi^0}Pb}$ are collected in the Fig. \ref{xie}. We can find that the results obtained from present model are deeper than that from the QMF model without pion and gluon corrections.

Single-$\Lambda$ and single-$\Xi^0$ energies of  these hypernuclei within QMF-NK3S set and the corresponding experimental data are listed detailedly in Table \ref{tab5}.  The differences of single-$\Lambda$ binding energies in theory and  experiment are less than $5\%$ of experimental values. The spin-orbit splittings of these single-$\Lambda$ hypernuclei are usually less than $0.3$ MeV, since the tensor couplings between vector meson and hyperons are included in this work. The small spin-orbit splittings are in accord with the available experiment data.  For single-$\Xi^0$ hypernuclei, the high angular momentum states do not exist comparing with the corresponding single-$\Lambda$ hypernuclei due to small value of single $\Xi$ potential at nuclear saturation density. In this case, the deepest bound state of $\Xi^0$ exists in $\ce{^{208}_{\Xi^0}Pb}$, about $-9.5$ MeV.

\begin{table}[htb]%\footnotesize%\scriptsize%
	\centering
	\caption{Energy levels (in MeV) of hyperon for $\ce{_Y^{40}Ca}$, $\ce{_Y^{89}Y}$, and $\ce{_Y^{208}Pb}$ by QMF-NK3S in the present model, compared with the experimental data.}
	\label{tab5}
	\begin{tabular}{l c c c c c c c c c}
		\hline
		\hline
		&$\ce{_\Lambda^{40}Ca}$(Exp.)&$\ce{_\Lambda^{40}Ca}$&$\ce{_\Xi^{40}Ca}$
		&$\ce{_\Lambda^{89}Y}$(Exp.)&$\ce{_\Lambda^{89}Y}$&$\ce{_\Xi^{89}Y}$
		&$\ce{_\Lambda^{208}Pb}$(Exp.)&$\ce{_\Lambda^{208}Pb}$&$\ce{_\Xi^{208}Pb}$\\
		\hline
		
		1$s_{1/2}$&$-18.7\pm1.1$&$-17.76$&$-5.22$&$-23.6\pm0.5$&$-22.64$&$-7.54$&$-26.9\pm0.8$&$-25.86$&$-9.41$\\
		
		1$p_{3/2}$&             &$-9.99$&$-1.32$&              &$-16.53$&$-4.11$&             &$-21.95$&$-6.99$\\
		
		1$p_{1/2}$&$-11.0\pm0.5$&$-9.78$&$-1.29$&$-17.7\pm0.6$&$-16.41$ &$-4.08$&$-22.5\pm0.6$&$-21.90$&$-6.98$\\
		
		1$d_{5/2}$&             &$-2.42$&       &             &$-9.88$ &$-0.60$&              &$-17.27$&$-4.18$\\
		
		1$d_{3/2}$&$-1.0\pm0.5$ &$-2.15$ &      &$-10.9\pm0.6$&$-9.65$ &$-0.56$&$-17.4\pm0.7$ &$-17.16$&$-4.15$\\
		
		1$f_{7/2}$&             &        &      &             &$-3.19$&        &              &$-12.10$&$-1.19$\\
		
		1$f_{5/2}$&             &        &      &$-3.7\pm0.6$ &$-2.89$&        &$-12.3\pm0.6$&$-11.92$&$-1.16$\\
		
		1$g_{9/2}$&             &        &      &             &       &        &             &$-6.66$&\\
		
		1$g_{7/2}$&             &        &      &             &       &        &$-7.2\pm0.6$ &$-6.40$&\\
		
		\hline
		\hline
	\end{tabular}
\end{table}

\begin{figure}[htb]
	\centering
	\includegraphics[bb=0 120 700 1400,width=7cm]{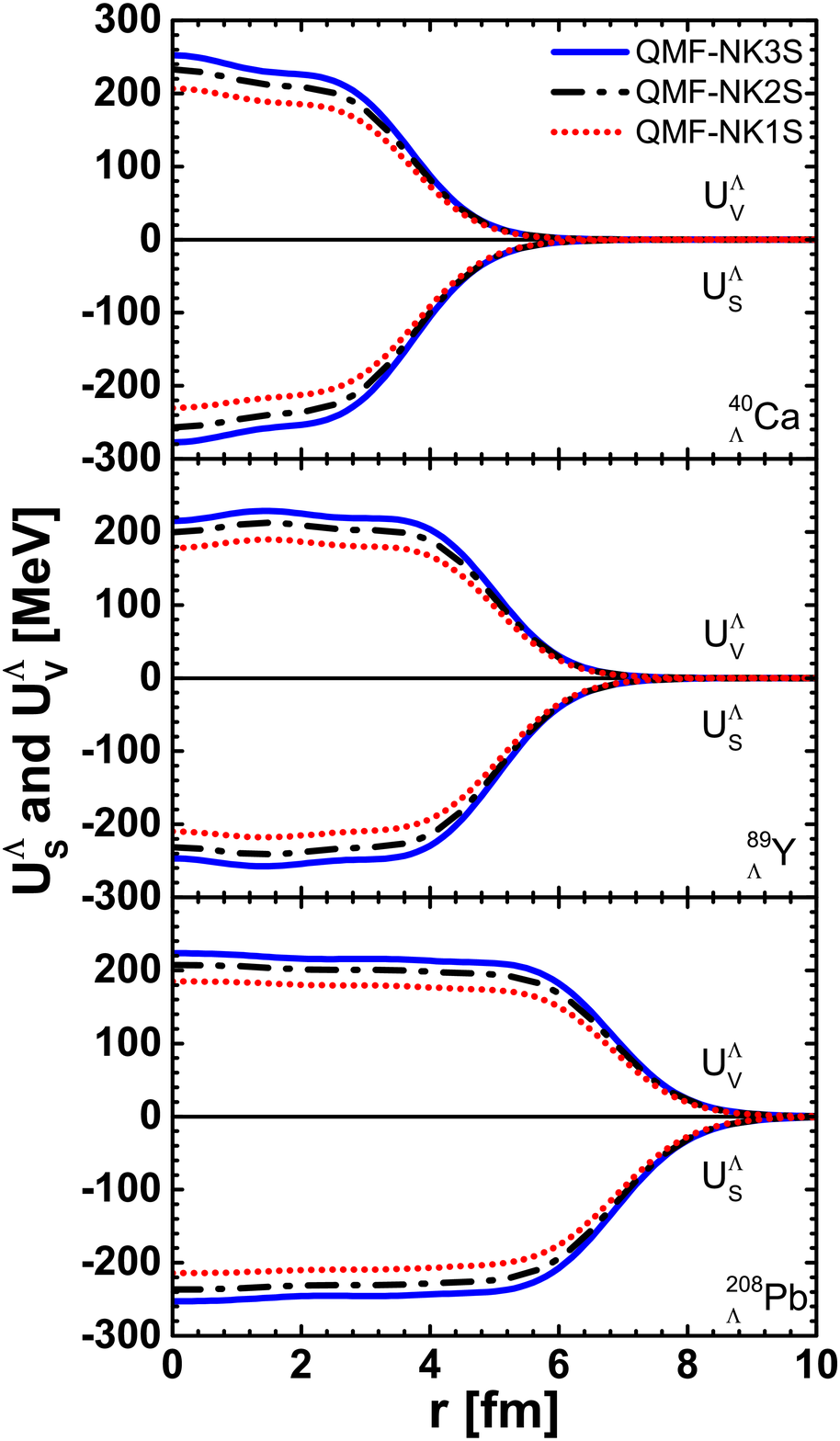}
	\caption{The scalar and vector potentials, $U_S^\Lambda$ and $U_V^\Lambda$, for $1s_{1/2}$ $\Lambda$ state in $\ce{_\Lambda^{40}Ca}$, $\ce{_\Lambda^{89}Y}$, and $\ce{_\Lambda^{208}Pb}$ by QMF-NK1S, QMF-NK2S, and QMF-NK3S.}
	\label{lpot}
\end{figure}

In Fig. \ref{lpot}, we plot scalar potential $U_S^\Lambda$ and vector potential $U_V^\Lambda$ for $1s_{1/2}$ $\Lambda$ state in $\ce{_\Lambda^{40}Ca}$, $\ce{_\Lambda^{89}Y}$, and $\ce{_\Lambda^{208}Pb}$. We can find that the scalar potential from $\sigma$ meson almost has the same magnitude as the repulsive vector potential from $\omega$ meson. They will cancel with each other and finally generate a total attractive force, whose center part is around 23-30 MeV in $\ce{_\Lambda^{40}Ca},~\ce{_\Lambda^{89}Y}$, and $\ce{_\Lambda^{208}Pb}$. The larger $u$ quark mass will provide more attractive scalar and repulsive vector potentials, which is related to smaller effective $\Lambda$ mass and larger vector coupling constant in QMF-NK3S parameter set.  

\begin{figure}[htb]
	\centering
	\includegraphics[bb=0 120 700 1400,width=7cm]{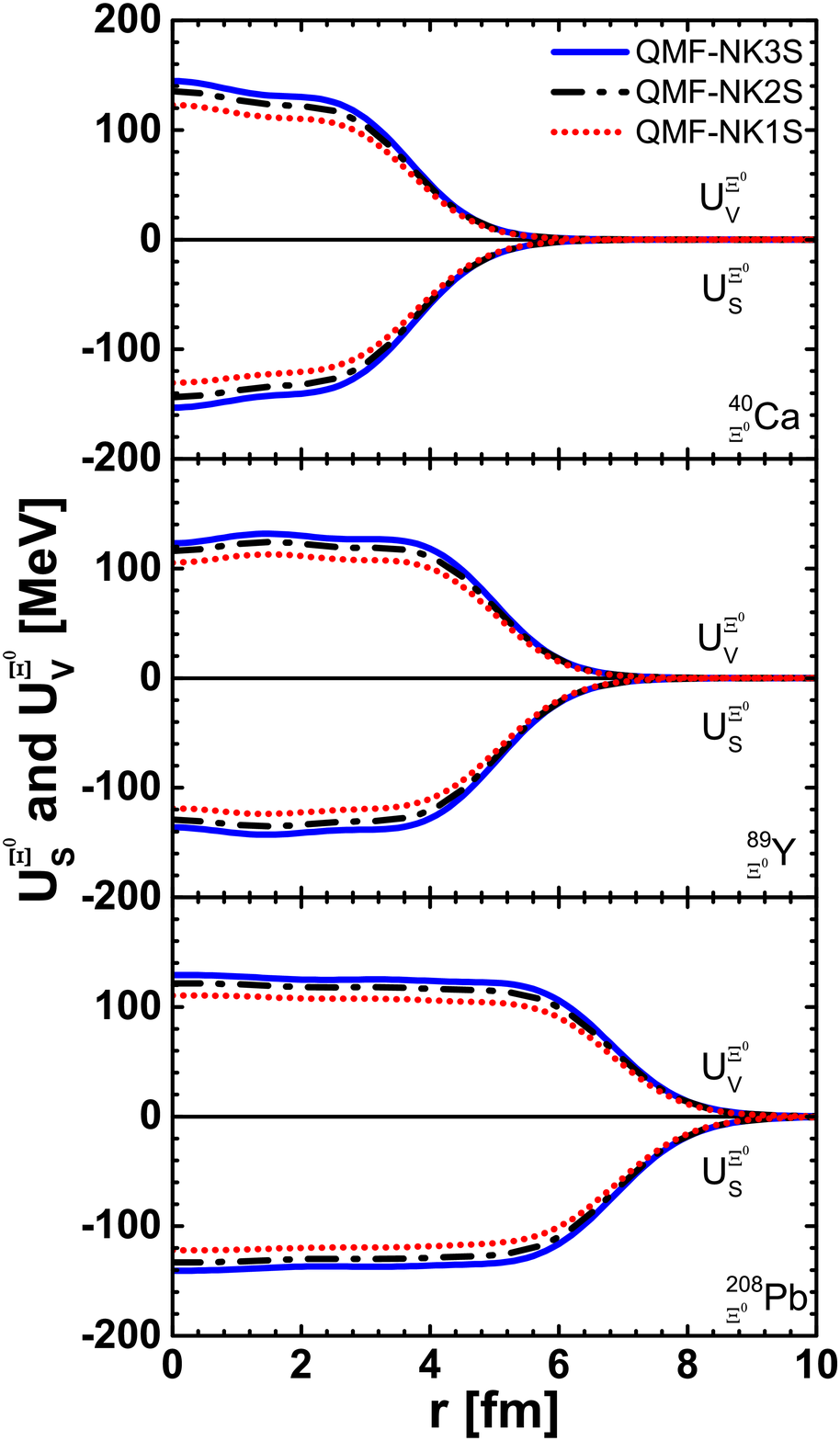}
	\caption{The scalar and vector potentials, $U_S^{\Xi^0}$ and $U_V^{\Xi^0}$, for $1s_{1/2}$ $\Xi^0$ state in $\ce{_{\Xi^0}^{40}Ca}$, $\ce{_{\Xi^0}^{89}Y}$, and $\ce{_{\Xi^0}^{208}Pb}$ by QMF-NK1S, QMF-NK2S, and QMF-NK3S.}
	\label{xpot}
\end{figure}

Similarly, the scalar and vector potentials of single-$\Xi^0$ hypernuclei $1s_{1/2}$ $\Xi^0$ state are given in Fig. \ref{xpot}. They are just about $50\%$ in $\Lambda$ hypernuclei. The scalar and vector potentials finally produce attractive potentials at the center part of $\Xi^0$ hypernuclei whose values are about 7-12 MeV in $\ce{_{\Xi^0}^{40}Ca}$, $\ce{_{\Xi^0}^{89}Y}$, and $\ce{_{\Xi^0}^{208}Pb}$.

\begin{figure}[htb]
\centering
\includegraphics[width=12cm]{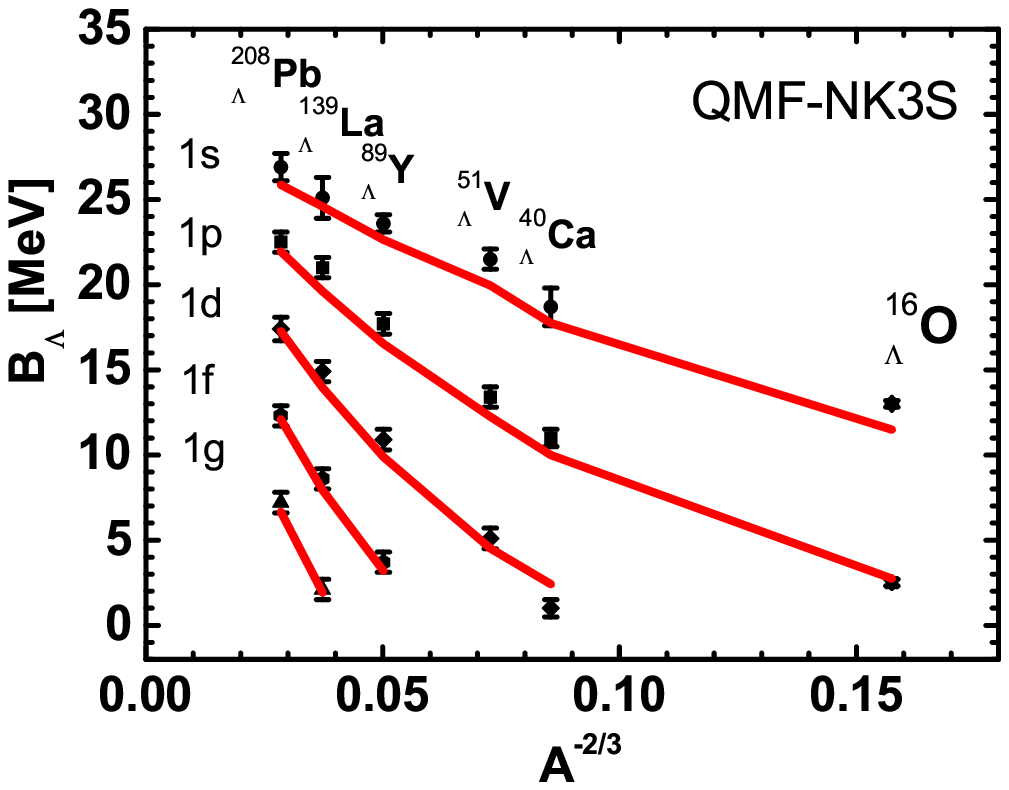}
\caption{Systematic calculations of the binding energies of  $\Lambda$ hypernuclei with QMF-NK3S parameter set compared with the experimental data.}
\label{bl}
\end{figure}

In Fig.~\ref{bl}, the binding energies of single $\Lambda$ hypernuclei are systematically calculated from  $\ce{_\Lambda^{16}O}$ to $\ce{_\Lambda^{208}Pb}$ within QMF-NK3S parameter set at different spin-orbit states and are compared with the experimental data~\cite{gal16}.  It can be found that the experiment observables are reproduced very well in QMF model including pion and gluon corrections. If pion and gluon corrections were not included~\cite{hu14a}, the $\Lambda$ binding energies at $s$ and $p$ spin-orbit states were in accord with experimental data, however, there were a few MeV differences of $\Lambda$ binding energies between theoretical results and experimental values above $d$ spin-orbit states.  It is convinced that the QMF model including the pion and gluon corrections can improve the description of $\Lambda$ hypernuclei from quark level.

\subsection{Properties of neutron stars}
\begin{figure}[htb]
	\centering
	\includegraphics[width=12cm]{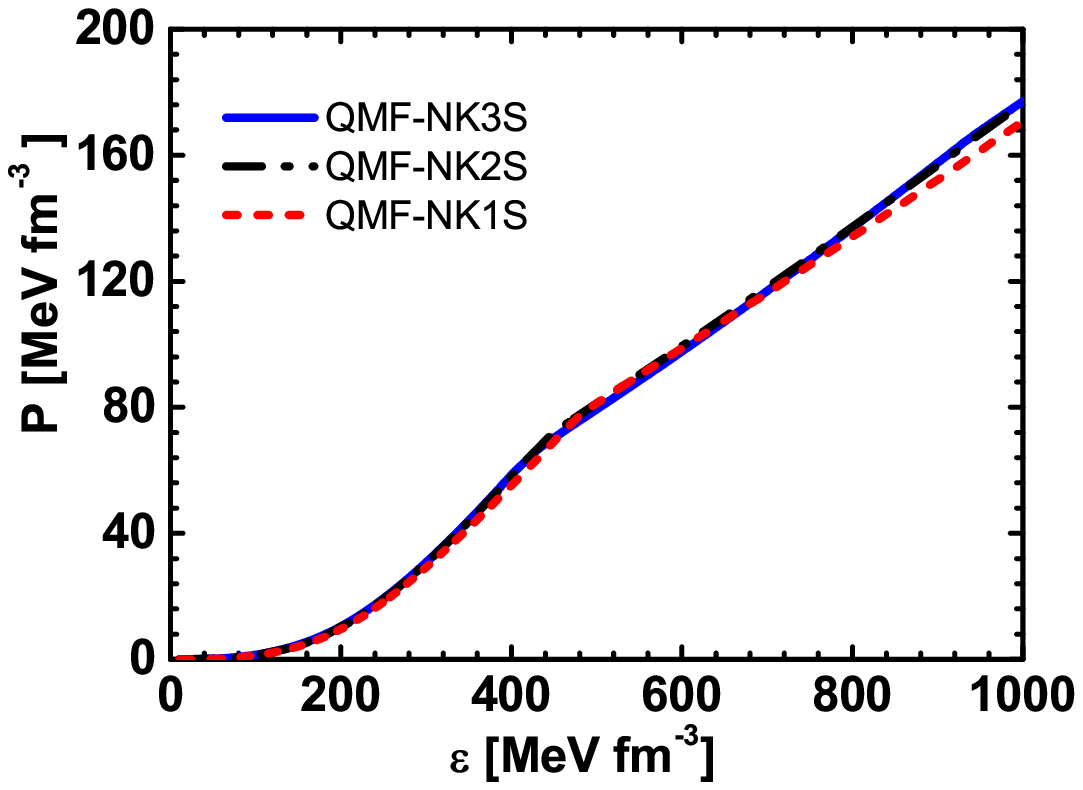}
	\caption{Pressures of $\beta$ equilibrated matter as  functions of the energy density, for cases QMF-NK1S, QMF-NK2S, and QMF-NK3S parameter sets.}
	\label{ep}
\end{figure}

Once the properties of single $\Lambda$ and $\Xi^0$ hypernuclei are drawn by the QMF model with pion and gluon corrections, the strange nuclear matter and neutron stars can be studied in present framework. For neutron star matter, all baryons and leptons stay in a charge neutrality and $\beta$ equilibrium environment. Furthermore,  $\Sigma$ hyperons also have the probability to appear in neutron star matter, although there is no evidence of existing of single $\Sigma$ hypernuclei in experiments until now. To make the present discussion simplify, the coupling constant between $\omega$ meson and $\Sigma$ hyperon is taken same value of $g^\Lambda_\omega$. Furthermore, the $\rho$ meson may play an important role in neutron star matter, whose coupling constants related to hyperons are chosen as $g^H_\rho=g_\rho$. After solving the corresponding equations, the energy density and pressure of neutron star matter can be obtained as shown in Fig. \ref{ep}  within QMF-NK1S, QMF-NK2S, and QMF-NK3S sets. At low energy density, the pressures of three parameter sets are almost identical, since the behaviors of neutron star matter at low density are decided by the properties of finite nuclei, meanwhile, the hyperons do not appear due to their larger chemical potentials. With energy density increasing, the pressure of QMF-NK1S becomes  a little bit of difference from those of QMF-NK2S and QMF-NK3S for hyperons appearance.  

\begin{figure}[htb]
	\centering
	\includegraphics[bb=0 100 330 550,width=7cm]{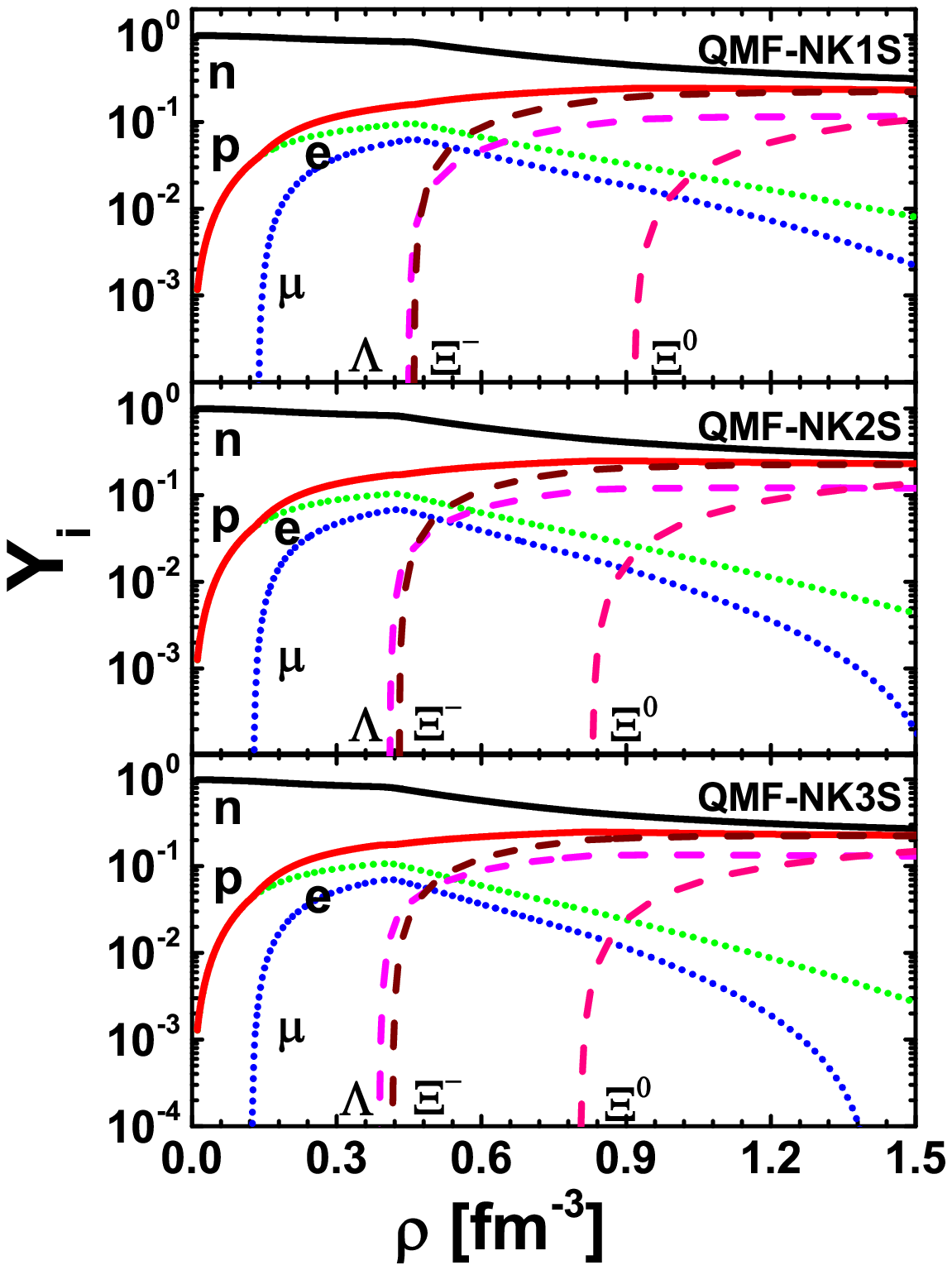}
	\caption{Fractions of leptons and baryons in neutron star matter as functions of total baryon density, for cases QMF-NK1S, QMF-NK2S, and QMF-NK3S parameter sets.}
	\label{yf}
\end{figure}

Besides the relation between energy density and pressure, the fractions of leptons and baryons in neutron star matter as functions of total baryon density are also given in Fig. \ref{yf} with different parameter sets. The direct Urca processes will happen above the densities $\rho_B=0.287, ~0.244$, and $0.229$ fm$^{-3}$ at QMF-NK1S, QMF-NK2S, and QMF-NK3S, respectively, which are higher than the case without pion and gluon corrections, $\rho_B=0.21$ fm$^{-3}$. It satisfies the constraint of astrophysical observations, where the cooling process does not occur at too low proton density.  Furthermore, both of $\Lambda$ and $\Xi^-$ hyperons appear around two times saturation density. $\Xi^0$ hyperons exist above $\rho_B=0.9$ fm$^{-3}$. At high density, the fraction of $\Xi^-$ hyperons approaches that of protons. Meanwhile, the fraction of $\Lambda$ hyperons is suppressed by $\Xi^-$ hyperons. Totally, the appearance of hyperons becomes earlier at larger $u$ quark mass.	 

\begin{figure}[htb]
	\centering
	\includegraphics[bb=0 30 290 210,width=14cm]{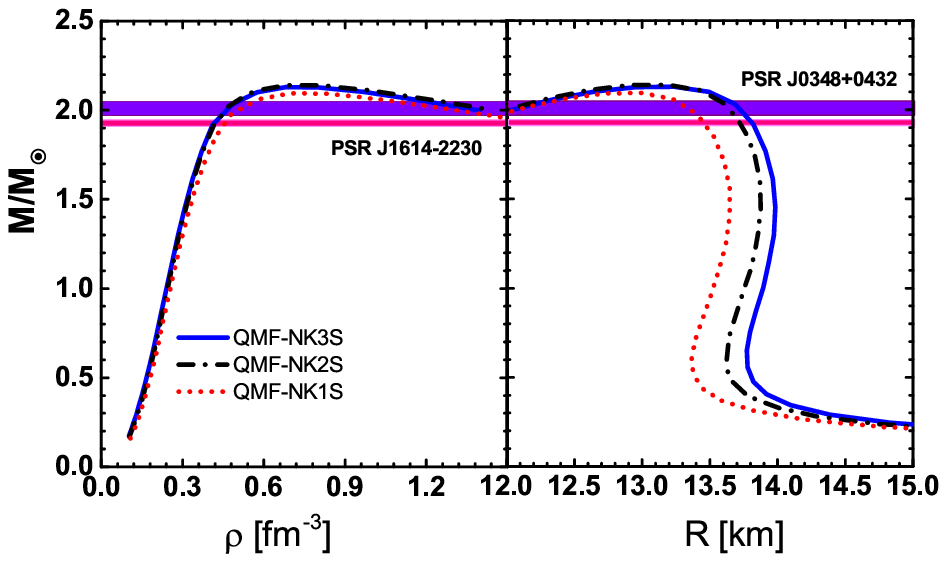}
	\caption{The masses of neutron stars as functions of density and radius, for QMF-NK1S, QMF-NK2S, and QMF-NK3S parameter sets.}
	\label{mr}
\end{figure}

By using the EOS of neutron star matter to solve TOV equation, the properties of neutron stars, such as the masses as functions of central density and radius, are obtained in Fig.~\ref{mr}. The maximum masses are $2.09$ to $2.14M_\odot$ generated by QMF-NK1S to QMF-NK3S, respectively. These results are in accord with recent astronomical observations of two massive neutron stars, PSR J1614-2230 ($1.928\pm0.017 M_\odot$)~\cite{demorest10,fonseca16} and PSR J0348+0432 ($2.01\pm0.04 M_\odot$)~\cite{antoniadis13}, moreover, the hyperons $\Lambda$ and $\Xi$ can exist in the core region of neutron star.  The densities corresponding to maximum masses are around $\rho_B=0.7$ fm$^{-3}$. The radii of these neutron stars distribute $13.0$ to $13.2$ km. They are larger than the constraint region worked by Hebeler {\it{et al.}}~\cite{hebeler10,hebeler13} around $11$ km, but smaller the values from MQMC framework. Actually, the radius of a neutron star still has not been measured directly until now. Comparing our previous work in  QMF model without pion and gluon corrections~\cite{hu14b}, the description to properties of neutron stars is largely improved to satisfy the constraint of observations in present QMF parameter sets. 

\section{Conclusion}
We have studied the properties of single $\Lambda, ~\Xi^0$ hypernuclei and neutron stars with hyperons in terms of quark mean field (QMF) model including pion and gluon corrections, where the baryons are composed of three independent relativistic quarks confined by a harmonic oscillator potential mixing with scalar and vector components. Corrections due to the center-of-mass motion, pionic and gluonic exchanges were considered in calculating properties of baryons perturbatively. The baryon-baryon interactions were generated by exchanging $\sigma,~\omega,$ and $\rho$ mesons between quarks at different baryons in a mean field approximation.

 The strengths of $s$ quark confinement potentials and constituent quark mass were determined by fitting free baryon masses of $\Lambda,~\Sigma^0,$ and $\Xi^0$ hyperons with a least-squares fitting method as a whole. The coupling constants between $u,~d$ quarks and mesons were already obtained in our previous work about normal nuclei.  Those between $s$ quark and mesons were decided through the potentials of $\Lambda N$ and $\Xi N$ at nuclear saturation density, as $U_\Lambda=-30$ MeV and  $U_\Xi=-12$ MeV, respectively.  Finally, we obtained three parameter sets corresponding to $u$ quark mass $m_u=250$ MeV, $m_u=300$ MeV, and $m_u=350$ MeV, named as QMF-NK1S, QMF-NK2S, and QMF-NK3S.

Energy levels of single $\Lambda$ hyperon for three hypernuclei, $\ce{_\Lambda^{40}Ca}$, $\ce{_\Lambda^{89}Y}$, and $\ce{_\Lambda^{208}Pb}$ were calculated. The results were consistent with the experiment observations very well and were largely improved comparing to those from previous QMF model without pion and gluon corrections, especially for high angular momentum states. Meanwhile,  energy levels of single $\Xi^0$ hyperon in $\ce{_{\Xi^0}^{40}Ca}$, $\ce{_{\Xi^0}^{89}Y}$, and $\ce{_{\Xi^0}^{208}Pb}$ were also obtained . The results for $\Xi^0$ hypernuclei could serve as a reference for the future experiments. The $\Lambda$ binding energies  from $\ce{_\Lambda^{16}O} $ to $\ce{_\Lambda^{208}Pb}$ were also compared systematically to the experimental data and in accord with them very well.

Finally, properties of neutron stars were studied in present framework. The coupling constants of $\Lambda$ and $\Xi$ hyperon were kept as the same values used in hypernuclei. The coupling constant between $\omega$ meson and $\Sigma$ hyperon was chosen the same value as $\Lambda$ hyperon. It was found that the $\Lambda$ and $\Xi^-$ hyperons started to appear in neutron star at two times nuclear saturation density and $\Xi^0$ hyperons at five times nuclear saturation density. The $\Sigma$ hyperons did not exist in the core of neutron star in this work. The maximum masses of neutron stars were around $2.09$ to $2.14 M_\odot$ within QMF-NK1S, QMF-NK2S, and QMF-NK3S sets, which satisfied the requirement of  recent astronomical observations about the massive neutron stars. The corresponding radii of neutron stars were about $13$ km.

The present QMF model including the pion and gluon corrections could describe properties of both hypernuclei and neutron stars, and satisfy the constraint of experimental data. In this work, to simplify our study, the charged $\Xi^-$ and double $\Lambda$ hypernuclei were not discussed, however many experiments are proposed to study their properties in large facilities. The related work is in progress. 

\section*{Acknowledgments}
This work was supported in part by the National Natural Science Foundation of China (Grant No. 11375089 and Grant No. 11405090) and the Fundamental Research Funds for the Central Universities.

\end{document}